\documentclass[]{spie}  

\usepackage{amsmath,amsfonts,amssymb}
\usepackage{subcaption}
\usepackage{graphicx}
\usepackage{cite} 
\usepackage{times}
\usepackage{epsfig}
\usepackage{amsmath}
\usepackage{nccmath}
\usepackage{amssymb}
\usepackage{mwe}
\usepackage{acro}
\usepackage{amssymb}
\usepackage{xcolor,colortbl}
\usepackage{tabularx}
\usepackage{relsize}
\usepackage{pifont}
\usepackage{booktabs} 
\usepackage{multirow}
\usepackage{multicol}
\usepackage{adjustbox}
\usepackage{float}
\usepackage{graphicx}
\usepackage{makecell}
\usepackage{tabu}
\usepackage[colorlinks=true, allcolors=blue]{hyperref}
\usepackage[capitalize]{cleveref}

\title{Multi-scale Multi-site Renal Microvascular Structures Segmentation for Whole Slide Imaging in Renal Pathology}

\author[*a]{Franklin Hu}
\author[*a]{Ruining Deng}
\author[b]{Shunxing Bao}
\author[c]{Haichun Yang}
\author[a,b,c]{Yuankai Huo}

\affil[a]{Department of Computer Science, Vanderbilt University, Nashville, TN, USA}
\affil[b]{Department of Electrical and Computer Engineering, Vanderbilt University, Nashville, TN, USA}
\affil[c]{Department of Pathology, Microbiology and Immunology, Vanderbilt University Medical Center, Nashville, TN, USA}

\authorinfo{*Franklin Hu and Ruining Deng contributed equally to this paper\\  Corresponding author: Yuankai Huo: E-mail: yuankai.huo@vanderbilt.edu}

\begin{document} 
\maketitle

\begin{abstract}
Segmentation of microvascular structures, such as arterioles, venules, and capillaries, from human kidney whole slide images (WSI) has become a focal point in renal pathology. Current manual segmentation techniques are time-consuming and not feasible for large-scale digital pathology images. While deep learning-based methods offer a solution for automatic segmentation, most suffer from a limitation: they are designed for and restricted to training on single-site, single-scale data. In this paper, we present Omni-Seg, a novel single dynamic network method that capitalizes on multi-site, multi-scale training data. Unique to our approach, we utilize partially labeled images, where only one tissue type is labeled per training image, to segment microvascular structures. We train a singular deep network using images from two datasets, HuBMAP and NEPTUNE, across different magnifications (40$\times$, 20$\times$, 10$\times$, and 5$\times$). Experimental results indicate that Omni-Seg outperforms in terms of both the Dice Similarity Coefficient (DSC) and Intersection over Union (IoU). Our proposed method provides renal pathologists with a powerful computational tool for the quantitative analysis of renal microvascular structures.

\end{abstract}


\section{INTRODUCTION}
\label{sec:intro}  
Recent years have witnessed a heightened interest among researchers and clinicians in the renal microvasculature, a pivotal component of the intricate human microvasculature~\cite{lowenstein2016rebirth}. This fascination arises from the renal microvasculature's key role in maintaining kidney function and its involvement in various kidney diseases' pathogenesis~\cite{krishnan2021microvascular}. Mapping this system is crucial for diagnosing kidney conditions like renal artery stenosis, aneurysms, and tumors, fostering timely detection and targeted interventions~\cite{tham1983renal}.

\begin{figure*}[t]
\begin{center}
\includegraphics[width=1\linewidth]{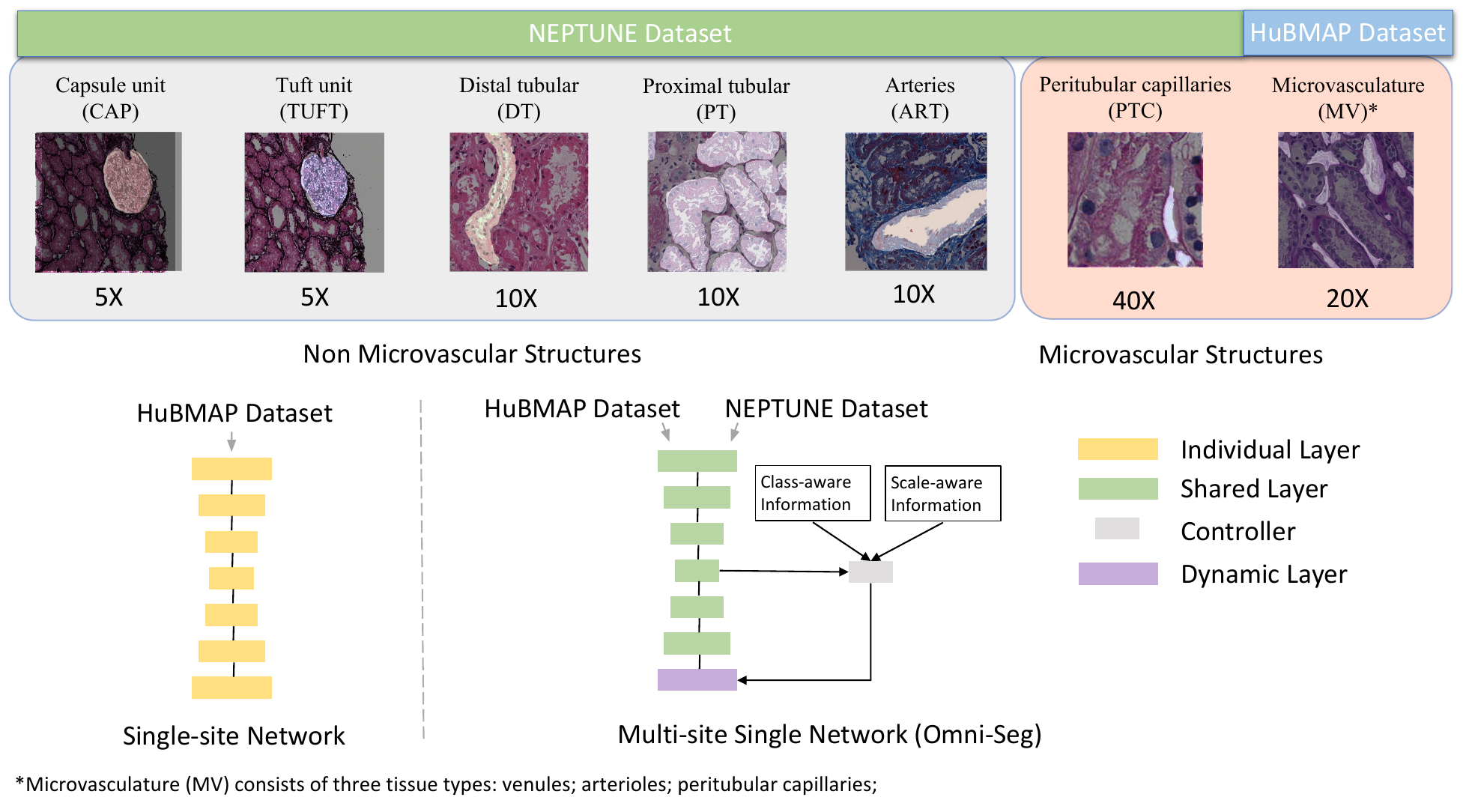}
\end{center}
\caption{Datasets from NEPTUNE and HuBMAP feature diverse partially labeled structures. While NEPTUNE contains various non-microvascular components, our strategy leverages this data to enhance model robustness. With the multi-site and multi-scale nature of these datasets, spanning magnifications from 5$\times$ to 40$\times$, our Omni-Seg method effectively addresses the challenge.}
\label{fig:problem}
\end{figure*}

Accurately quantifying renal microvasculature is a growing need in the scientific and medical sectors. This intricate blood vessel network within kidneys is instrumental in regulating blood pressure and preserving renal function~\cite{chade2017small}. An efficient tool for renal microvasculature measurement would not only bolster early detection and monitoring of kidney diseases but also promote the development of tailored treatments. Discussions about harnessing AI and deep learning for renal structure segmentation and quantification are gaining traction~\cite{huo2021ai, kannan2019segmentation}.

Recognizing the importance, The Human Biomolecular Atlas Program (HuBMAP) initiated the ``HuBMAP – Hacking the Human Vasculature" challenge, which ran from May 22 to July 31, 2023. This challenge, hosted on the acclaimed Kaggle platform, spurred participants to design robust algorithms for segmenting microvasculature instances in human kidney histology images~\cite{hubmap-hacking-the-human-vasculature}. Although HuBMAP offered 2D-PAS stain tissue samples across the renal landscape, relying on single-site data in machine learning poses risks. These models often face challenges with generalization, tending to overfit to specific site characteristics, compromising adaptability and robustness. Moreover, there's inherent uncertainty in their real-world performance due to potential biases.

To overcome these challenges, we integrated digital renal biopsies from the Nephrotic Syndrome Study Network (NEPTUNE)\cite{barisoni2013digital} with the HuBMAP data, creating a robust multi-site dataset. However, NEPTUNE's biopsies spanned various scales (5$\times$ to 40$\times$), contrasting HuBMAP's consistent 20$\times$ magnification. To meld the two datasets harmoniously, we deployed Omni-Seg\cite{deng2021omni}, a single dynamic network adept at managing multi-scale and multi-label data.

In this paper, we showcase an innovative approach leveraging multi-site multi-scale training data for segmenting microvascular structures in human kidney WSIs. Our objective is to arm renal pathologists with quantitative tools for understanding renal microvascular structures, enriching diagnostic and research insights.

Our proposed method makes a remarkable contribution in two key aspects: (1) To the best of our knowledge, Omni-Seg represents the first attempt at a multi-site, multi-scale network specifically designed for microvascular segmentation. Notably, it accomplishes this by effectively utilizing multiple datasets that span various scales. (2) The segmentation network we propose exhibits considerable potential, achieving higher segmentation performance scores when compared to other existing approaches. This success paves the way for more accurate and reliable segmentation of kidney whole slide images, enabling more precise analysis and interpretation of microvascular structures in renal pathology.

\begin{figure*}[t]
\begin{center}
\includegraphics[width=1\linewidth]{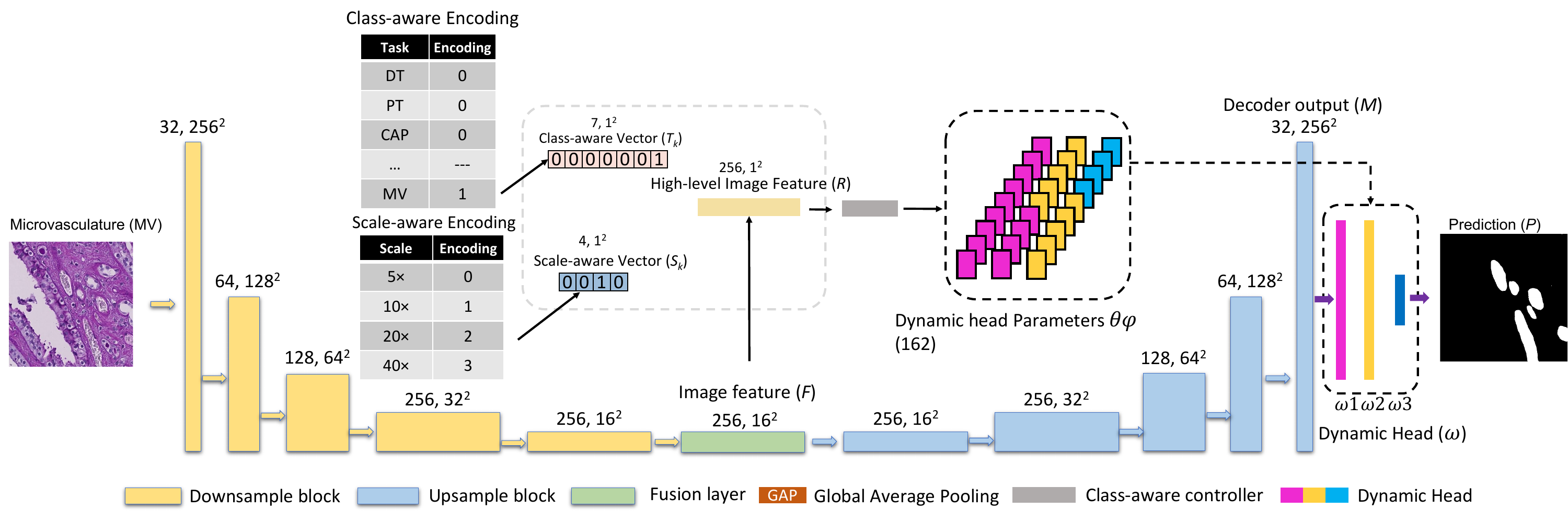}
\end{center}
\caption{This figure illustrates the Omni-Seg pipeline, which consists of three primary components: a residual U-Net backbone, a class-aware controller, and a dynamic segmentation head. To augment performance, we've integrated a class-aware knowledge encoder, aiding in the acquisition of domain-specific insights from the multi-label dataset. This holistic strategy empowers the model to adeptly learn and adapt to the data's diverse features, resulting in enhanced segmentation outcomes.}
\label{fig:pipeline}
\end{figure*}

\section{Method}

\subsection{Dynamic multi-label modeling}
In partially labeled datasets~\cite{jayapandian2021development}, each training image is labeled for only one tissue type. This characteristic makes these datasets unsuitable for the direct training of standard multilabel segmentation networks. To overcome this challenge, our proposed method introduces a class-aware and scale-aware encoding approach. The knowledge associated with different tissue types is represented as an $m$-dimensional one-hot vector, where $m$ denotes the total number of tissue types. Meanwhile, the knowledge associated with different scales is represented as an $n$-dimensional one-hot vector, where $n$ denotes the total number of magnifications for the pathological image. This encoding strategy effectively leverages the partial labeling information and scale knowledge during training~\cite{chen2017fast}. The formula for this encoding process is provided in the subsequent equation:

\begin{center}
\begin{equation}
T_k =
\begin{cases}
    1, & if ~~ k = i \\
    0, & otherwise
\end{cases}
\quad \text{for } k = 1, 2, \ldots, \emph{m}
\end{equation}
\end{center}

\begin{center}
\begin{equation}
S_p =
\begin{cases}
    1, & if ~~ p = j \\
    0, & otherwise
\end{cases}
\quad \text{for } p = 1, 2, \ldots, \emph{n}
\end{equation}
\end{center}

To integrate domain-specific information with the embedded features, the class-aware vector ($T_k$) and the scale-aware vector in $p$-th scale ($S_p$) are reshaped to align with the shape of the image features. The image feature, denoted as $F$, undergoes global average pooling (GAP), yielding a feature vector in $\mathbb{R}^{N \times 256 \times 1 \times 1}$, where $N$ signifies the batch size. The high-level image features are then concatenated with the reshaped class-aware vector. Subsequent to this step, a single 2D convolutional layer, known as the class-aware controller $\phi$, is utilized as a feature-based fusion block. This operation refines the fused features, forming the final controller for the dynamic head mapping process, as described in the following equation.

\begin{center}
\begin{equation}
\omega = \phi(\text{GAP}(F)||T_k||S_p; \Theta_{\phi})
\end{equation}
\label{fusionequ}
\end{center}

Our input is formulated from the fusion of feature $GAP(F)$, the class-aware vector $T_k$, and the scale-aware vector $S_p$ using the fusion process $||$. The size of the parameters $\Theta_{\phi}$ for the domain-aware controller dictates the output channel and serves as the foundation for determining the dynamic head's parameters. Consequently, our task encoding and scale encoding technique exhibits lower computational and spatial complexity when contrasted with the multi-network approach.

\subsection{Dynamic head mapping}
Drawing inspiration from DoDNet~\cite{zhang2021dodnet}, our methodology employs a binary segmentation network fortified with a dynamic filter to facilitate multi-scale segmentation. Through multi-label modeling, we derive joint low-dimensional image feature vectors, class-aware vectors, and scale-aware vectors at the ideal segmentation magnification. These elements are subsequently harnessed to govern a streamlined dynamic head, which discerns the specific tissue type and resolution information from the input.

The dynamic head is comprised of three layers: the initial two layers feature eight channels, whereas the concluding layer showcases two channels, accumulating a sum of 162 parameters. We directly correlate the parameters from the fusion-based feature controller with the kernels present in the dynamic head, thus enabling meticulous segmentation steered by multi-modality features. The mechanism of this filtering procedure can be explicated by Eq.~\ref{dynamichead}.

\begin{equation}
\begin{split}
 	    P = ((((M * \omega_1) * \omega_2) * \omega_3)
\end{split}
\label{dynamichead}
\end{equation}

In the above, $*$ signifies convolution, $M$ denotes the output from the decoder, while $\omega$ represents the dynamic heads. The expression $P$ in $\mathbb{R}^{N \times 2 \times W \times H}$ provides the prediction, where $N$, $W$, and $H$ respectively correspond to the batch size, width, and height of the dataset.

\subsection{Residual U-Net backbone}
The segmentation backbone employed in our approach is grounded on the residual U-Net, as depicted in DoDNet~\cite{zhang2021dodnet} (refer to Fig.~\hyperref[fig:pipeline]{2}). Although DoDNet was configured around a 3D network design, we adapted it for 2D pathological images by integrating 2D convolutional blocks with a kernel size of $3 \times 3$. Following each convolutional block, a ReLU activation is employed, succeeded by group normalization~\cite{wu2018group}.

To extract high-level image features, we introduced a convolutional fusion layer equipped with a $3 \times 3$ kernel. Downsampling blocks with a stride of 2 were deployed to halve the dimensions of the input feature map. As a result, the feature maps underwent adjustments via multiple encoder-decoder blocks spanning different pyramid scales.

In a parallel manner, the decoder enlarges the feature maps using an upsample factor of 2, while concurrently reducing the channel count by half. Within each upsample block, a low-level feature map (sourced from the analogous encoder layer) merges with an upsampled feature map, undergoing refinement via a residual block. This feature-centric learning mechanism culminates in the generation of high-level features pivotal for segmentation.

\subsection{Testing on ground truth labels}
During the testing interval, addressing the discrepancy in image dimensions between the NEPTUNE and HuBMAP datasets emerged as a focal point. While NEPTUNE images spanned $256 \times 256$ pixels, the HuBMAP images were more expansive, extending to $512 \times 512$ pixels. To achieve uniformity and guarantee model compatibility, a preprocessing maneuver was devised for the NEPTUNE images, wherein four distinct $256 \times 256$ images were amalgamated to produce a singular $512 \times 512$ image. By embracing this methodology, we were able to preserve a harmonized input template across both datasets, thereby streamlining their incorporation into the model for thorough evaluation and scrutiny.

\section{Experiments}

\subsection{Data}

For the NEPTUNE dataset, 1751 regions of interest (ROIs) were captured from 459 WSIs, originating from 125 patients diagnosed with minimal change diseases. These images underwent manual segmentation for six structurally normal pathological primitives: glomerular tuft (TUFT); glomerular unit (CAP); proximal tubular (PT); distal tubular (DT); peritubular capillaries (PTC); and arteries (ART) as detailed in~\cite{jayapandian2021development}. The segmentation used digital renal biopsies sourced from a multi-center Nephrotic Syndrome Study Network (NEPTUNE)~\cite{barisoni2013digital}. Originally, these images possessed dimensions of 3000$\times$3000 pixels at a 40$\times$ magnification and were stained using Hematoxylin and Eosin (HE), Periodic-acid-Schiff (PAS), Silver (SIL), and Trichrome (TRI). These staining methods functioned as color augmentations for each tissue type. Subsequently, the images were cropped and downsampled to 256$\times$256 patches for optimal magnification~\cite{jayapandian2021development}. Although the dataset was initially divided into a 6:1:3 ratio for training, validation, and testing, the absence of fully segmented images of microvascular structures precluded us from leveraging the testing set. Instead, our model was trained using only the training and validation sets, incorporating the PTC data into our microvascular (MV) class.

Complementing the NEPTUNE dataset, we also incorporated data from HuBMAP. This dataset is comprised of 5 PAS-stained WSIs from varied donors, chosen based on criteria such as image quality (minimal artifacts or blurring), demographic diversity (considering age, sex, BMI), and encompassing different kidney regions (cortical, medullary, papillary). Expert segmentation was performed on the WSIs using QuPath by a lead anatomist, assisted by four other trained anatomists. They identified three microvascular structures: arterioles, venules, and capillaries. These were later grouped under a single category termed ``microvasculature"~\cite{hubmap-hacking-the-human-vasculature}. The WSIs were then transformed into 5440 images of dimensions 512$\times$512 at a 20$\times$ magnification. This dataset was sectioned into a 3:1:1 ratio for training, validation, and testing. The testing segment provided us with 1088 images, which served as the basis for our result analysis.

\begin{figure*}[t]
\begin{center}
\includegraphics[width=1\linewidth]{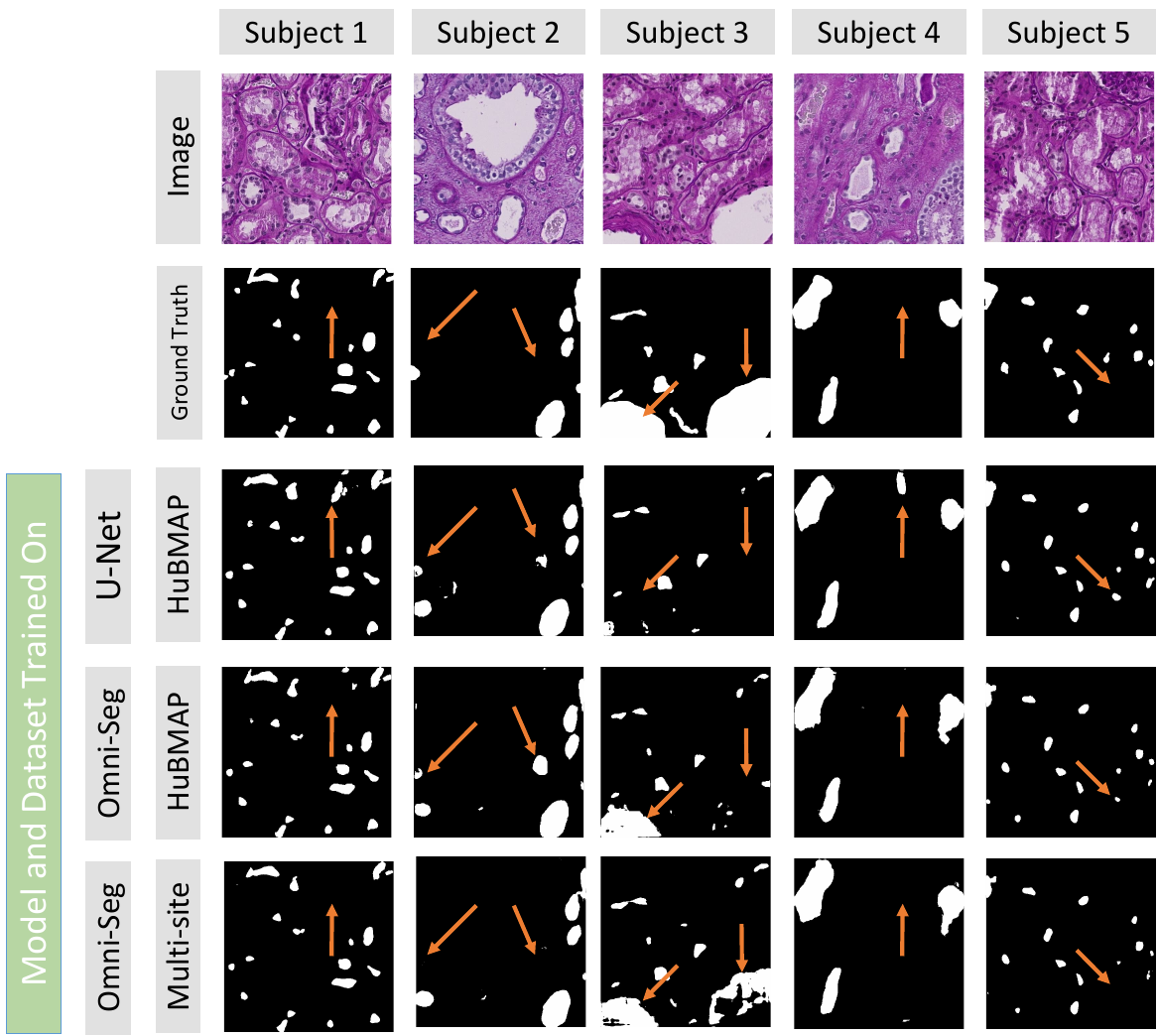}
\end{center}
\caption{The figure provides a visual representation of the qualitative results obtained from various methods. Notably, brown arrows highlight areas of interest and differences between the methods.}
\label{qualitative}
\end{figure*}

\subsection{Experimental details}
During the training process, seven image pools were created to organize batches containing patches from the same tissue type. This approach drew inspiration from the Cycle-GAN image pool\cite{zhu2019multi}. The batch size was set to four, and each image pool could hold up to eight images. When the pool size exceeded the batch size, images from the pool were fed into the network.

For loss calculation, we employed a combination of Binary Dice loss and cross-entropy loss during each backpropagation for all tasks. To emphasize accurate boundary prediction, a higher weight of 1.2 was assigned to the boundaries of the ground truth. This technique is prevalent in segmentation tasks\cite{ronneberger2015u}. Weight updates were carried out using Stochastic Gradient Descent (SGD) with a learning rate of 0.001 and a decay rate of 0.99.

Data augmentation techniques, such as Affine transformations, Flips, Contrast adjustments, Brightness adjustments, Coarse Dropouts, Gaussian Blurs, and Gaussian Noise, were applied to the entire training dataset across all methods. These augmentations were implemented using the imagaug package~\cite{jung2019imgaug} with a probability of 0.5.

For performance evaluation, we used two primary metrics: Dice Similarity Coefficient (DSC) and Intersection over Union (IoU). The mean DSC served as the main criterion for selecting the optimal model in the validation set. The model's performance was subsequently evaluated using the testing set. To maintain consistency, we selected checkpoints with the highest scores achieved across 100 epochs of training and validation.

All experiments were executed on a workstation boasting 32 GiB of RAM and an NVIDIA RTX A5000 GPU. This setup ensured a standardized computational environment for reliable and reproducible results.


\begin{center}
\label{table}
Table 1: Model performance on testing set (\%)
\begin{tabular}{c c c c c} 
 \hline
Model/Data & Median DSC & Mean DSC  & Std Dev DSC & Mean IoU \\
 \hline
 U-Net/HuBMAP  & 81.29 & 79.86 & \textbf{9.33} & 46.61 \\ 
Omni-Seg/HuBMAP     & 82.09 & 79.78  & 9.75 & 46.64 \\
Omni-Seg/Multi-site   & \textbf{82.30} & \textbf{80.26} & 9.36 & \textbf{47.46} \\
 \hline
\end{tabular}
\end{center}

\section{Results}

In this study, we conducted a comparative analysis of the proposed Omni-Seg pipeline, trained on multi-site data, against two alternative methods. The first method is the single-site U-Net pipeline~\cite{ronneberger2015u}, and the second is the Omni-Seg model trained exclusively on the HuBMAP dataset. Fig.~\hyperref[qualitative]{3} highlights the qualitative differences among the predictions generated by these three methods.

Several key distinctions arise from our comparisons. Firstly, both the U-Net with HuBMAP data and the HuBMAP-trained Omni-Seg methods produce predictions that misalign with the ground truth. Conversely, our multi-site Omni-Seg approach exhibits better alignment with the actual microvascular structures. Moreover, the multi-site trained Omni-Seg model excels at identifying areas within the ground truth that the other methods either miss or fail to recognize as microvascular structures.

These findings underscore the advantages of leveraging multi-site data in the Omni-Seg pipeline. This leads to more accurate and robust microvascular segmentation in human kidney whole slide images. The results emphasize the potential benefits of integrating diverse datasets and highlight the superior performance of the multi-site Omni-Seg method in detecting essential microvascular structures for renal pathology analysis.

Table~\hyperref[table]{1} displays consistent results across all datasets, with our Omni-Seg/Multi-site method achieving marginally higher mean and median dice scores than the other two methods. The standardized testing set from the single HuBMAP data source substantiates these findings.

While all three methods exhibit significant capabilities in predicting microvasculature (with median dice scores exceeding 0.8), our method emerges as the top performer, boasting the superior mean DSC, median DSC, and mean IoU scores. This emphasizes the efficacy of our Omni-Seg/Multi-site approach in precisely segmenting microvascular structures in human kidney whole slide images.

The uniform performance across diverse datasets accentuates the robustness and generalizability of our method, establishing it as a dependable option for exhaustive microvascular segmentation tasks in renal pathology.

\section{Conclusion}
In this paper, we introduce Omni-Seg, a dynamic segmentation network tailored for multi-site, multi-scale segmentation of renal microvasculature. By harnessing both microvascular and non-microvascular images, our proposed method achieves high-performance segmentation of renal microvasculature. This strategy exhibits immense potential as an automated tool, delivering accurate and efficient quantification of renal microvascular structures for renal pathologists.

\section{ACKNOWLEDGMENTS}
This work has not been submitted for publication or presentation elsewhere. It is supported in part by NIH R01DK135597(Huo), DoD HT9425-23-1-0003(HCY), and NIH NIDDK DK56942(ABF).

\bibliography{main} 
\bibliographystyle{spiebib} 

\end{document}